\documentclass[preprint,preprintnumbers,amsmath,amssymb]{revtex4}

\usepackage{graphicx}
\usepackage{amssymb}
\usepackage{inputenc}

\begin{document}

\title{Strange quark matter fragmentation in astrophysical events}

\author{L. Paulucci}
\email{laura.paulucci@ufabc.edu.br}
\affiliation{Universidade Federal do ABC \\
Rua Santa Ad\'elia, 166, 09210-170 Santo Andr\'e, SP, Brazil}

\author{J. E. Horvath}
\affiliation{Instituto de Astronomia, Geof\'\i sica e Ci\^encias
Atmosf\'ericas - Universidade de S\~ao Paulo\\
Rua do Mat\~ao, 1226, 05508-900, Cidade Universit\'aria, S\~ao Paulo SP, Brazil}

%\date{}

\begin{abstract}
 The conjecture of Bodmer-Witten-Terazawa suggesting a form of quark matter
(Strange Quark Matter) as the ground state of hadronic interactions
has been studied in laboratory and astrophysical contexts by a large number
of authors. If strange stars exist, some violent events involving these
compact objects, such as mergers and even their formation process,
might eject some strange matter into the interstellar medium that could be
detected as a trace signal in the cosmic ray flux. To evaluate this possibility,
it is necessary to understand how this matter in bulk would fragment in the
form of {\it strangelets} (small lumps of strange quark matter in which
finite effects become important). 
We calculate the mass distribution outcome using the 
statistical multifragmentation model and point out
several caveats affecting it. In particular, the possibility
that strangelets fragmentation will render a tiny fraction of contamination in
the cosmic ray flux is discussed.

\bigskip

{\bf keywords:} strange quark matter, strangelets, statistical multifragmentation model

\end{abstract}

\maketitle

\section{Introduction}
Sometime after the papers of Bodmer \cite{Bodmer} and Terazawa \cite{Terazawa} put forward
the idea of a quark ground state of strongly interacting matter, the wide and colorful
discussion given by Witten \cite{Witten} added considerable interest to the issue of what
is now called the {\it Strange Quark Matter} (SQM) hypothesis.

The stability scenario has been systematically studied for the first time by Farhi and 
Jaffe \cite{Farhi} within the MIT bag model, where a wide parameter space for absolute 
stability to hold was established. More recently, it has been claimed that the preferred 
state would be when quarks form pairs, similarly to electrons in ordinary superconductivity, 
for it would allow an even lower energy per baryon number for the system due to the 
formation of the color condensate \cite{pairing1, pairing2, pairing3}.

If the SQM hypothesis is valid, the low probability of a {\it simultaneous} decay of roughly 
a third of up and down quarks in a nuclei into strange quarks under everyday conditions 
would prevent the transition. However, it has been shown \cite{trans1, trans2, trans3} 
that for nuclear systems at high density and moderate temperature, the transition could be 
favored. In this way, compact objects are naturally though as niches for the existence of SQM. 
Among the predicted systems and phenomena, strange stars \cite{SS1, SS2, SS3, SS4}, 
compact stars where the transition to SQM happens in all the stellar interior, and the 
possibility of strangelets \cite{Farhi, Madsen98, Madsen02}, small lumps of strange 
quark matter, were discussed. The possibility of strangelets being a part of the cosmic
ray flux, and likely involved in exotic events \cite{Bjorken}, naturally raised the
question about the conditions for bulk strange quark matter to break apart, and with what
mass and energy spectra the fragmentation into strangelets would ensue \cite{HV}.

Starting with these early papers, some injection mechanisms for strangelets in astrophysical
sites have been proposed: strange stars mergers \cite{Witten, Rosinska}, phase 
transition during type II supernovae \cite{Mac1, Mac2, HV}, and acceleration in strange 
pulsar environment \cite{ChengUsov}. All these process might lead to a measurable abundance 
of this component among the cosmic ray flux, although they have not been addressed in 
full detail yet. Considering this, a simple manner of testing the existence
of strange matter in the interior of compact stars would be the detection by ground-based or
in-orbit experiments of strangelets of astrophysical origin. In fact, some experiments claimed
to have detected possible exotic components \cite{Data1, Data2, Data3, Data4}, though 
a live debate has taken place without a firm confirmation of the nature of the primaries.

On the other hand, the problem of fragmentation of nuclear matter during cooling/decompression
has been studied for several decades (for a review, see for example \cite{Bondorf}), 
with wide applicability to laboratory experiments (e.g., nuclei collision in accelerators). 
A range of r-process nuclei could be produced this way \cite{Schramm} and the
analogue situation with SQM in the place of nuclear matter appears to be justified if the
Bodmer-Witten-Terazawa conjecture is true.
However, the production of strangelets and subsequent acceleration still lack a detailed
general analysis. In particular, the predicted fluxes of strangelets among cosmic rays are
based on plausible suppositions instead of refined calculations, and in general dismiss
the possible decay of these particles into ordinary nuclei \cite{Madsen05}.
Moreover, the energy spectrum of strangelets to be injected in the interstellar medium is
quite uncertain. In this way, supposing the Fermi mechanism will accelerate strangelets
the same way it does with ordinary cosmic rays may pose some problems \cite{Medina},
since only particles with a non-thermal spectrum can be accelerated by shocks. The
distribution of masses and energies at the injection site are then important ingredients for all
these attempts to connect some events with the possible strangelet primaries arriving to earth \cite{Rosswog}.

Here we present an analysis for the fragmentation of strange quark matter within
the statistical multifragmentation model. This model can be applied to a supernova
explosion driven by the conversion of ordinary nuclear matter to strange quark matter,
for example, being this scenario an alternative to the neutrino-driven ones that still face
difficulties in explaining the explosion in numerical simulations. The conversion
during the proto-neutron star phase could provide enough energy for the expel of
stellar material either in the form of a detonation wave or of a second neutrino
wind \cite{Mac1, Mac2}. The ejected outer layers could be contaminated by
strangelets due to turbulent mixing effects \cite{mix}. As a general
result, we shall show that a fragmentation into mass chunks having $A \leq 100$ may be
expected, although significant uncertainties in the underlying physics remain, and in 
fact recent calculations do not obtain ejection of SQM \cite{HR}. Since the 
temperatures and other parameters are quite similar, scenarios of the merging 
of two strange stars \cite{Rosswog} would follow a similar fragmentation 
pattern.
 
In a recent paper, Biswas and collaborators \cite{Biswas} used the statistical
multifragmentation model to analyze the fragmentation of strange quark matter in a
scenario of strange stars mergers, concluding that the mass spectrum results in low mass fragments and shows an
exponential decay with $A$ and also presenting an estimate for the strangelet 
flux based on cosmic ray diffusion
properties. However, their analysis dismissed important contributions to the energy
of the fragments and assumed some physical properties
(discussed at length in \cite{tese}) that may significantly alter the results.
These are related to the dependency of the energy of the fragment on the
strange quark mass (assumed negligible)
and also on the possibility of pairing between quarks, as
will be described in the next section.

\section{Statistical Multifragmentation Model}

Among the several proposals formulated to deal with the fragmentation problem,
the statistical multifragmentation model (SMM)
(see \cite{Bondorf} and references therein) has provided consistent results when applied to
the bulk nuclear matter $\rightarrow$ nuclei transition.

When we proposed using the SMM to treat the fragmentation of strange
quark matter \cite{NICIX}, we initially employed it to treat fragmentation
in a supernova driven by the conversion of nuclear matter into SQM scenario.
Recent works have shown that in the collision of two strange stars, matter 
achieve high temperatures \cite{T-estimate} (of order $\sim$ tens of $MeV$),
which are high enough to comply with the hypotheses of the statistical 
multifragmentation model. Specifically, the critical condition involving
the excitation energy per baryon number for the occurrence of the
break-up (which must be comparable to the total binding energy,
ensuring thermal and dynamical equilibrium) is satisfied. Therefore, in 
both scenarios the fragmentation should proceed similarly. Also, in ref. 
\cite{NICIX}, we used an approximate treatment for the strangelet
energy taking mean values instead of considering the full dependence of the 
surface and curvature energies on temperature and baryonic number, 
which is certainly important for the matter, and considered strange quark 
matter without pairing. We found in ref. \cite{NICIX} some inconsistencies 
regarding the position of the fragmentation peak, as we shall discuss bellow,
the present treatment is the result of this analysis.

We based our analysis on a simplified version of the SMM \cite{Bugaev00} in which the
system is studied in the grand canonical ensemble, rendering neat analytical solutions when
the thermodynamical limit is taken. Generally speaking, an exponential behavior for
the partition function is predicted for high masses.

We have started from the partition function of a single fragment with $A$ nucleons

\begin{equation}\label{omega}
\omega_A=V\Big(\frac{mTA}{2 \pi}\Big)^{3/2}e^{-f_A/T},
\end{equation}
\\
where $f_A$ is the internal free energy of the fragment

\begin{equation}\label{free_energy}
f_A=-WA+\sigma A^{2/3}+CA^{1/3},
\end{equation}
\\
and $W$ represents the volume binding energy per baryon number of SQM, 
$\sigma$ and $C$ being the internal free energy of a fragment with baryon number $A$, 
rest mass $m$ and chemical potential $\mu$ corresponding to the surface
and curvature contributions, respectively; and $T$ is the bulk SQM temperature.

From the definition of pressure in the grand canonical ensemble,

\begin{equation}
p(T,\mu)=T\lim_{V \to \infty}\frac{\ln \mathcal{Z}(V,T,\mu)}{V},
\end{equation}
\\
where $\mathcal{Z}$ is the Laplace transform of the grand canonical partition
function, the pressures for both phases are obtained from the singularities of
the isobaric partition function (for details, see \cite{Bugaev00} and references therein).

The liquid and gas pressures are given by

\begin{eqnarray}
p_g(T,\mu)&=&T\Big(\frac{mT}{2\pi}\Big)^{3/2}\Big\{z_1e^{\frac{\mu-bp_g}{T}}+\sum_{A=2}^{\infty}A^{3/2}e^{[(\nu-bp_g)A-\sigma A^{2/3}-CA^{1/3}]/T}\Big\}, \\
p_l(T,\mu)&=&\frac{\nu}{b},
\end{eqnarray}
\\
where $\nu=\mu+W$ is the (shifted) chemical potential.

The fragmentation spectrum, $\mathcal{P}_g$, can be then obtained (considering chemical equilibrium
between bulk matter and the fragments) by taking the derivative of the gas pressure, $p_g$,
with respect to the chemical potential of the fragments, $\mu_A$,

\begin{equation}\label{pg}
\mathcal{P}_g(A)=\frac{\partial}{\partial\mu_A}p_g =
\Big(\frac{m_0T}{2\pi}\Big)^{3/2}A^{3/2}e^{\big[(\mu+W-bp_g)A-\sigma
A^{2/3}-CA^{1/3}\big]/T},
\end{equation}
\\
In the model, the parameter $b$ represents the repulsive interactions in
a simple Van der Waals approximation.

We have considered strange quark matter within the MIT bag model framework, in the
color-flavor-locked (CFL) state \cite{pairing1, pairing2, Lugones}. The energy of
each fragment was calculated by employing the multiple reflexion 
expansion formalism as in \cite{SQMT}, thus presenting the necessary
dependence on the temperature, baryonic number, gap parameter, bag constant, 
and strange quark mass. 

When obtaining the mass number for which the fragment distribution reaches 
its maximum in the coexistence region, we have checked that the peak is 
always obtained for strangelets with mass numbers $A \ll 1$, leaving only
the exponential behavior being apparent. 
It is not clear what is the meaning of this result.
One possibility is that the description is actually incomplete, or
one cannot use the grand canonical
approach either. An alternative interpretation is that the system would
{\it not} fragment at all, remaining in the bulk SQM state in spite of the 
mechanical and thermal perturbations to which it is subject in the outbreak 
and further expansion.
In particular, thermal equilibrium between the bulk and the
fragments throughout the whole process had to be assumed in the calculation and may
not be valid. If the system is to fragment in few large chunks of matter, then 
an statistical approach would not be adequate. 

However, one factor which can be important is related to the
presence of the vacuum, represented by the bag constant $B$. This term is
naturally absent when considering nuclear matter fragmentation since the
parameters obtained for describing it already consider the influence of the
vacuum. But here it is possible to look at the fragmentation
of SQM as a process in which a fraction of the vacuum energy is used to
provide strangelets with surface and curvature energies among other finite size effects.
Therefore, there is a difference of energy density per baryon number between the
liquid and gaseous phases which should be taken into account. 

In this way, we 
have introduced the bag constant directly into the energy density of the gas 
and liquid (bulk) phases by substituting the volume internal free energy per 
baryon number for $W=W_0+Bv$ and we shall continue to use this approach throughout
the rest of our analysis. This last term is not the same for both
phases due to the dependency with the proper volume associated with each system:

\begin{eqnarray*}
W_l=W_0+Bv_{liq}, \\
W_g=W_0+Bv_{gas}.
\end{eqnarray*}

The density of fragments with baryon number $A$ is given by Eq. \ref{pg}.
The argument of the exponential in the mass distribution $[(\nu-bp_g)A]$ in
the coexistence phase ($p_g^*=p_l$) is now

\begin{eqnarray*}
[\nu-bp_g^*]A&=&[\nu_g-\nu^*_l\,]A=[\mu_g+W_g-\mu_l-W_l\,]A\\
&=&[B\,(v_{gas}-v_{liq})\,]A,
\end{eqnarray*}
\\
where $v$ is the volume per baryon number.

\begin{figure}
\begin{center}
\includegraphics[width=0.49\textwidth]{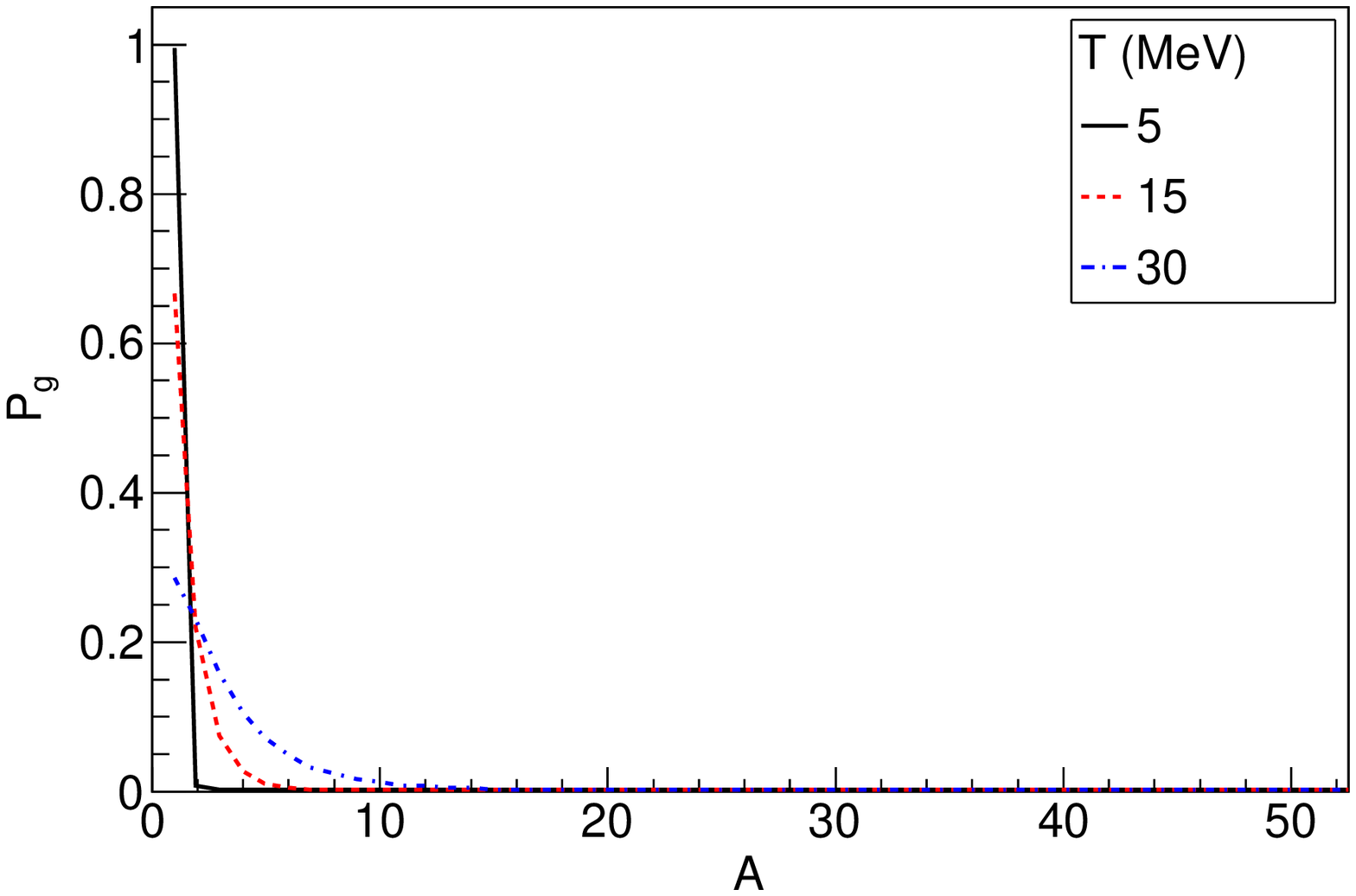}
\includegraphics[width=0.49\textwidth]{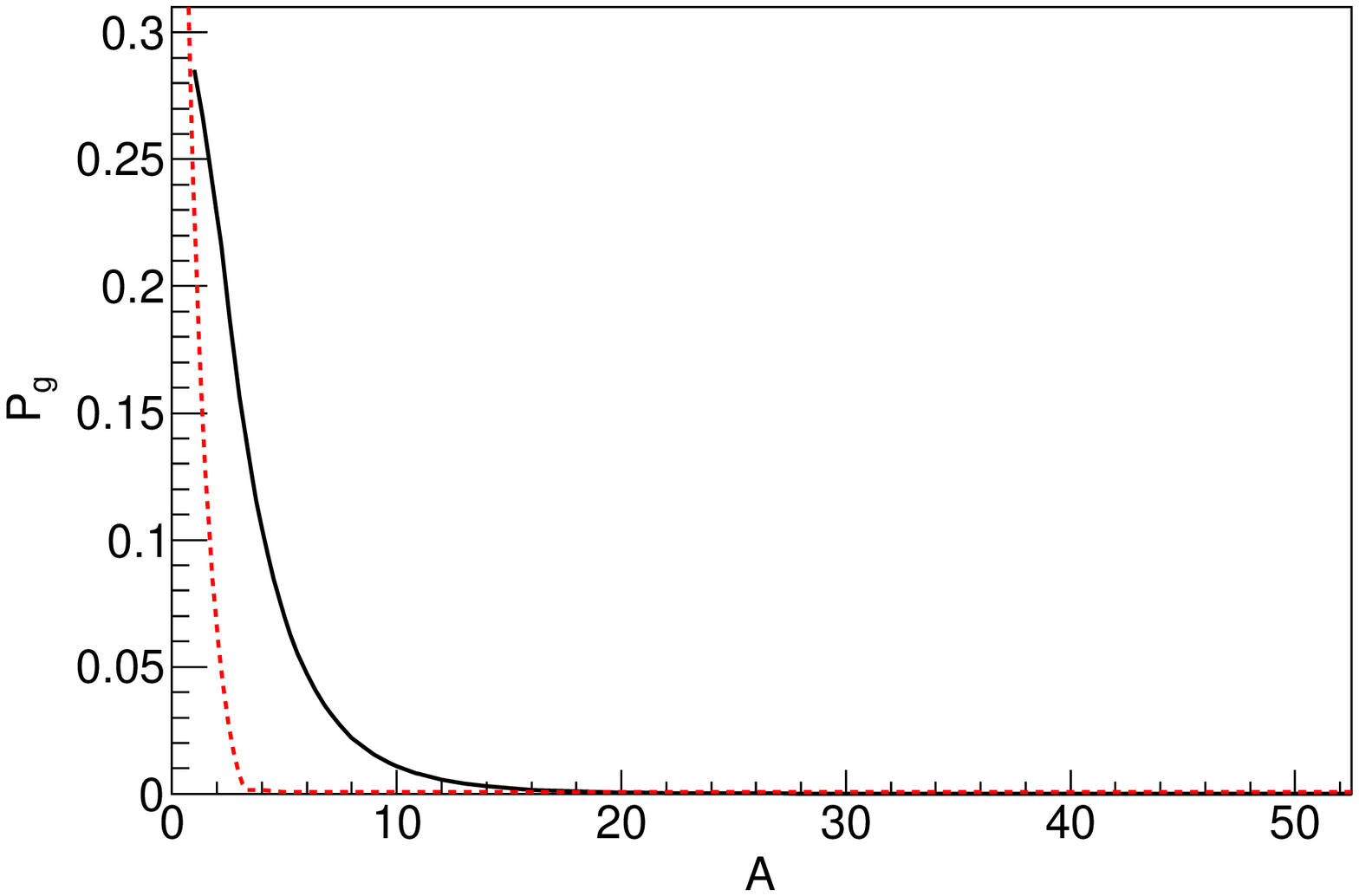}
\caption{\footnotesize Normalized distribution function of fragments for CFL
strangelets with $B^{1/4}=145$ MeV, $m_s=100$ MeV, and $\Delta=50$ MeV. On the
left panel, we show its dependence on the temperature considering the SMM with 
direct introduction of the bag constant. On the right panel, a comparison of 
the results obtained for a given temperature with (full line) and without 
(dashed line) the introduction of the vacuum energy.}\label{pico2}
\end{center}
\end{figure}

Following the approach for deriving the temperature dependent internal free energy
presented in \cite{SQMT}, we have obtained the normalized  mass distribution function shown in
Figure \ref{pico2}.

We see that the whole fragment distribution is now shifted to higher
values of A, although the peak is still not in a physical position.
Also, we notice that although increasing the system's temperature
leads to a less stable system, it also decreases the values of
the surface and curvature's terms \cite{SQMT}, thus favoring strangelets with
higher $A$. It must be pointed out, nevertheless, that high temperature
strangelets would be more prone to evaporation \cite{Witten} and would have to
cool down in order to survive.

\begin{figure}
\begin{center}
\includegraphics[width=0.49\textwidth]{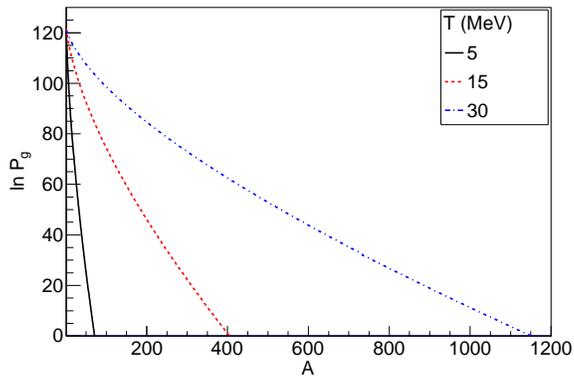}
\caption{\footnotesize Distribution function of fragments for CFL
strangelets with $B^{1/4}=145$ MeV, $m_s=100$ MeV, and $\Delta=50$ MeV
considering the ejection of $10^{-4} M_{\odot}$ as a function of the 
temperature.}\label{lnpg}
\end{center}
\end{figure}

\begin{figure}
\begin{center}
\includegraphics[width=0.49\textwidth]{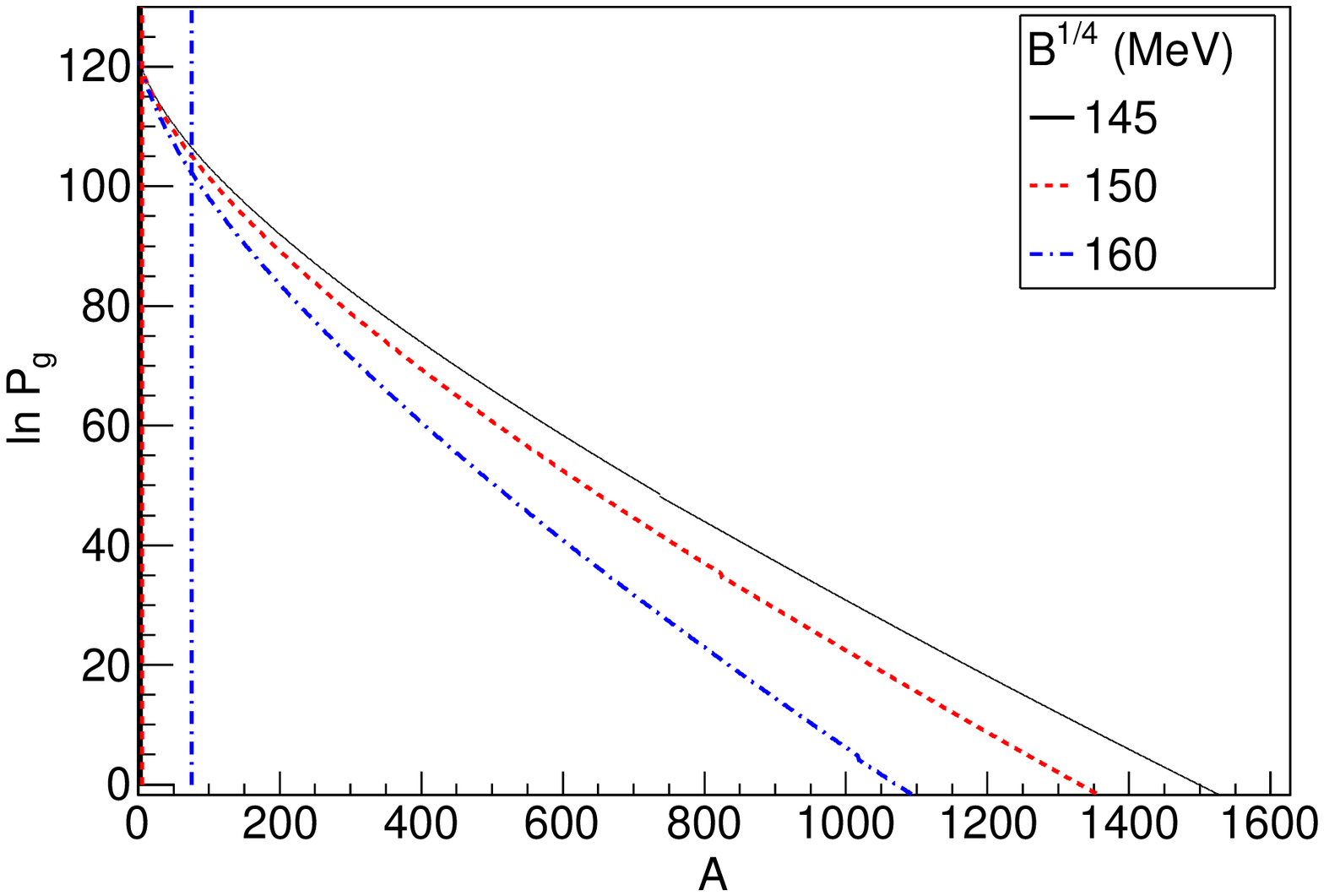}
\includegraphics[width=0.49\textwidth]{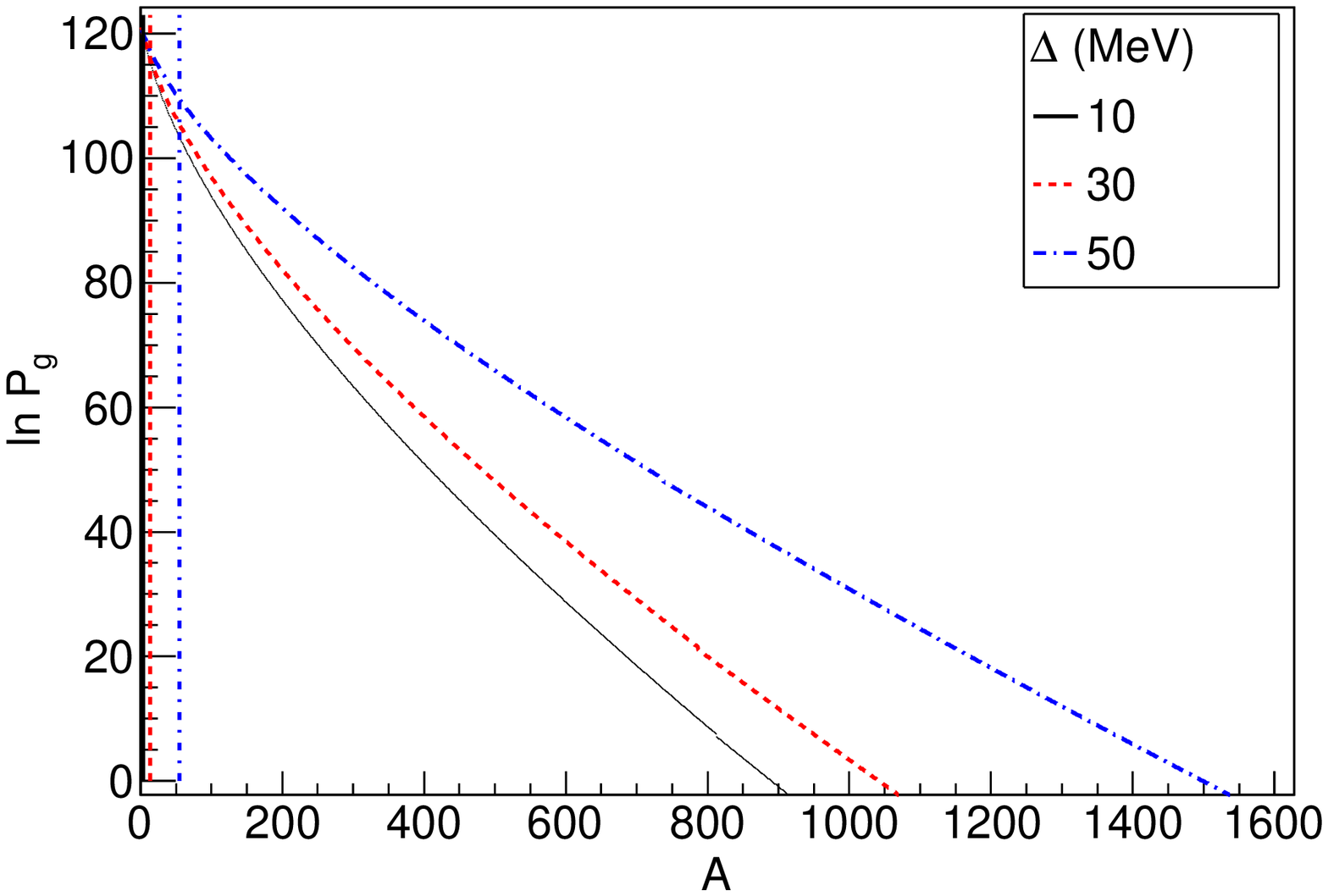}
\caption{\footnotesize On the left, distribution function of fragments for CFL
strangelets with $T=30$ MeV, $m_s=150$ MeV, and $\Delta=50$ MeV
considering the ejection of $10^{-4} M_{\odot}$ as a function of value of the bag
constant. The stability limit, i. e., the value of $A$ below which strangelets 
are not stable, is represented by the vertical lines.
On the right, the same but fixing the bag constant at $B^{1/4}=145$ MeV
and varying $\Delta$.}\label{lnpg_BD}
\end{center}
\end{figure}

Figure \ref{lnpg} considers a strangelet injection scenario with
the possible ejection of $10^{-4} M_{\odot}$ so we
can compare our results with the one presented in reference \cite{Biswas}.
For the values presented of strange quark mass, bag constant and gap
parameter, strange quark matter would be stable for almost all masses
(the minimum baryon number for stability is $A=4$). Nevertheless, we
stress that this form of SQM is obtained using almost ``optimal''
values for these unknown parameters, and therefore may not be very realistic. 
Any increase in $B$ or $m_s$ or
a decrease in $\Delta$ would lead to a much less stable strange matter,
this instability being greater as the temperature increases, as can
be seen in Figure \ref{lnpg_BD}.

\begin{figure}
\begin{center}
\includegraphics[width=0.49\textwidth]{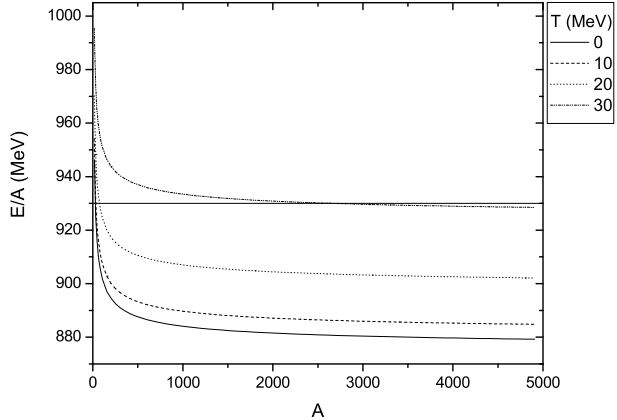}
\caption{\footnotesize Energy of strangelets without pairing with 
$B^{1/4}=145$ MeV, $m_s=150$ MeV as a function of $A$ for a fixed  
temperature, as indicated. The horizontal line indicates the threshold for 
strangelets to decay to normal nuclear matter.}\label{ENSQM}
\end{center}
\end{figure}

It is also important to remark that if the superconducting phase 
is not considered, strangelets would be
even more unstable. For example, for $B^{1/4}=145$ MeV and $m_s=150$ MeV,
at zero temperature, strangelets with $A \lesssim 15$ would be unstable but
at 30 MeV, only those with $A \gtrsim 2700$ would not decay to normal nuclear 
matter as is exemplified in Figure \ref{ENSQM}.
In this way, in this scenario of bulk strange quark matter fragmentation 
driven by expansion,
the existence of a large fraction of strangelets in the cosmic
ray flux is highly unlikely and would certainly be negligible if color
superconductivity is not considered, as already
proposed in reference \cite{Medina}. Also, if all quarks were assumed
to have zero mass, the stability of this system would be artificially enhanced
since the surface tension is associated with a non-zero strange quark mass.
Both simplified featured were employed in reference \cite{Biswas}. These 
remarks explain why in the work of Biswas and collaborators is claimed that
the amount of light fragments is increased with temperature with the suppression
of heavy fragments, in opposition of what is seeing in this work.

It has been suggested \cite{Alford08}, however, that strange stars may
present a strangelet crust embedded in a electron background. If this is the
case, then during a merger event strangelets would already be ejected with a 
mass spectrum with baryonic number of a few hundreds and one should expect 
a considerable amount of strangelets in the cosmic ray flux, an idea so far not 
favored by experiments.

\section{Conclusions}

We have presented an analysis of the fragmentation of SQM into strangelets in
high-temperature astrophysical settings. We have tried to make explicit the
assumptions made to calculate the distribution of fragments and how the
uncertainties could affect the final result.

We note that previous studies have argued that SQM must fragment into 
very large pieces \cite{Madsen02},
comparable to asteroid sizes, when mechanically stressed by external fields.
The physical argument was that, in
opposition to nuclear matter, the energy per baryon number always decreases with an
increase in $A$ for SQM. This indicates that it is necessary to introduce a large
amount of energy external to the system to break SQM into smaller (but very macroscopic)
fragments. This conclusion is not based on the application of a fragmentation
framework, and even if true it is clear that it would not emerge from an
analysis within the grand canonical ensemble.

We have shown that the simplest fragmentation of bulk SQM into strangelets yield an
odd result within the SMM, since the fragment peak would fall in an
unphysical region, perhaps in agreement with the previous expectations \cite{Madsen02}: matter would remain
in the bulk phase provided the SMM model consistently describes the process.
Alternatively, one can question the very validity of the latter for this
description. It has also been shown that the explicit consideration of
the vacuum energy density shifts the fragmentation spectra towards higher values
of $A$, although the position of the peak is still for $A \ll 1$. This tentative
mode for SQM fragmentation results in strangelets with a characteristic mass
scale as shown in Fig. (\ref{lnpg}). This points to a possible negligible
presence of this exotic particles in the cosmic ray flux.

The work of Biswas and collaborators \cite{Biswas} disregarded an important
feature in the SQM energy derivation, the fact that the strange quark mass
is of order of 100 MeV \cite{Davies}, giving rise to a significant surface energy.
It will certainly increase the total energy of the system with obvious implications
to the minimum baryon number for absolute stability to hold. Also, the
inclusion of an extra term in the exponential in Eq. \ref{pg} will make
the fragment distribution decrease with a faster pace with increasing $A$. In this way, most strangelets
fragmenting in their scenario would decay into normal nuclear matter as soon as
formed.

Our own results can be interpreted as indicating that most of the SQM will not fragment 
and that mostly unstable fragments (those subject to evaporation at a given temperature) 
will dominate the process, leading to the production of ordinary clusters and nucleons. 
Moreover, when considering higher temperatures, the spectrum is shown to extend 
to higher masses (Fig. \ref{lnpg_BD}), but the strangelets are more vulnerable to evaporation 
and the net outcome is no strangelets at all. Therefore, we conclude that a 
non-negligible amount of strangelets could only result from milder temperatures and 
paired quarks. Even if a huge fraction of SQM remains in bulk it will be subject to 
evaporation and should be affected, possibly surviving if the pairing is strong enough 
\cite{LugHor}.

The attempts made by us to confirm/validate the mass fragmentation spectra of SQM have rendered
ambiguous and/or inconsistent results. For instance, the minimization of
information entropy of Aichelin \& Huefner
\cite{Aichelin} which, in principle, could be adequate for the
merging of strange stars and the supernova ejection alike, predicts,
in general, a peak in the mass distribution, but its extension to large macroscopic
masses is difficult and the results obtained inconclusive. The same is true
when one tries a very simple approach of constructing the phase region {\`a la Gibbs},
that is, using the conditions $T_{gas}=T_{liq}$, $\mu_{gas}=\mu_{liq}$, and $P_{gas}=P_{liq}$. It is precisely the question of whether or not there would be a large number of fragments that could be treated in an statistical manner that is behind these ambiguities.

If the process is such that one can use the
grand canonical formalism for its study and take the solution to the thermodynamic
limit, or if a random sampling directly dealing with the microcanonical ensemble of
all the decay channels is necessary, or if this process can happen out of equilibrium
(violating the Gibbs criteria) are still open questions.
There is a final overall {\it caveat} concerning the role of the residual strong interaction
between nucleons, since it has a different behavior at large distances than
gluon-exchange forces mediating interactions
between quarks. The crude approximations built-in in any of these formalisms may hide its
true importance and reliability of the results themselves. This is an important
subject deserving a deeper understanding in order to provide better predictions
of the possible contamination of strangelets in the cosmic ray flux.

\section*{Acknowledgments}
The authors wish to acknowledge the financial support received from
Funda\c c\~ao de Amparo \`a Pesquisa do Estado de S\~ao Paulo and
from the CNPq Agency (Brazil).

%% References
%%
%% Following citation commands can be used in the body text:
%% Usage of \cite is as follows:
%%   \cite{key}         ==>>  [#]
%%   \cite[chap. 2]{key} ==>> [#, chap. 2]
%%

%% References with BibTeX database:
\bibliographystyle{elsarticle-num}
\bibliography{fragmentation.bib}

\end{document}